\documentclass[final,1p,times]{elsarticle}

\usepackage{graphicx}

\usepackage{amssymb}

\journal{Physica C}

\begin{document}

\begin{frontmatter}



\title{Resonant soft x-ray scattering, stripe order, and the electron spectral function in cuprates}


\author{Peter Abbamonte}
  \address{Department of Physics and Frederick Seitz Materials Research Laboratory, University of Illinois, Urbana, IL, 61801, USA}
  \address{Advanced Photon Source, Argonne National Laboratory, Argonne, IL, 60439, USA}
\author{Eugene Demler}
  \address{Department of Physics, Harvard University, Cambridge, MA, 02138, USA}
\author{J. C. Seamus Davis}
  \address{Laboratory for Atomic and Solid State Physics, Department of Physics, Cornell University, Ithaca, NY 14853, USA}
  \address{Condensed Matter Physics and Materials Science Department, Brookhaven National Laboratory, Upton, NY 11973, USA}
  \address{School of Physics and Astronomy, University of St. Andrews, North Haugh, St. Andrews, Fife KY16 9SS, UK}
\author{Juan-Carlos Campuzano}
  \address{Materials Science Division, Argonne National Laboratory, Argonne, IL, 60439}
  \address{Department of Physics, University of Illinois, Chicago, IL, 60607, USA}

\address{}

\begin{abstract}
We review the current state of efforts to use resonant soft x-ray scattering (RSXS), which is an elastic, momentum-resolved, valence band probe of strongly correlated electron systems, to study stripe-like phenomena in copper-oxide superconductors and related materials.  We review the historical progress including RSXS studies of Wigner crystallization in spin ladder materials, stripe order in 214-phase nickelates, 214-phase cuprates, and other systems.

One of the major outstanding issues in RSXS concerns its relationship to more established valence band probes, namely angle-resolved photoemission (ARPES) and scanning tunneling microscopy (STM).  These techniques are widely understood as measuring a one-electron spectral function, yet a relationship between RSXS and a spectral function has so far been unclear.  Using physical arguments that apply at the oxygen $K$ edge, we show that RSXS measures the square modulus of an advanced version of the Green's function measured with STM.  This indicates that, despite being a momentum space probe, RSXS is more closely related to STM than to ARPES techniques.

Finally, we close with some discussion of the most promising future directions for RSXS.  We will argue that the most promising area lies in high magnetic field studies, particularly of edge states in strongly correlated heterostructures, and the vortex state in superconducting cuprates, where RSXS may clarify the anomalous periodicities observed in recent quantum oscillation experiments.
\end{abstract}


\end{frontmatter}


\section{Introduction}
One of the central challenges in the field of high temperature superconductivity is to understand how highly correlated, Mott-like insulators such as La$_2$CuO$_4$, YBa$_2$Cu$_3$O$_6$, etc., transform into high temperature superconductors upon the addition of holes.  The so-called ``underdoped" regime intervening the insulator and the superconductor is marked by frustrated competition between kinetic energy effects, i.e., the desire of holes to delocalize, and the desire of the system to retain local, antiferromagnetic order.\cite{kivelsonNature,kivelsonRMP,kivelsonPNAS}  How superconductivity emerges from this highly correlated environment is still a mystery.

There is widespread evidence that these systems can exhibit another instability -- distinct from superconductivity -- often referred to as the static stripe or ``stripe smectic" phase.\cite{kivelsonNature}  This phase exhibits spontaneous breaking of both translational and rotational symmetry in the form of quasi-long-ranged, coexisting spin and charge order.  From a symmetry point of view, there is no distinction between this phase and a conventional density wave instability, such as the Peierls charge density wave (CDW) in NbSe$_3$ or the spin density wave (SDW) in elemental Cr, both of which are driven by a divergence in the static susceptibility arising from Fermi surface nesting. An important distinction, however, is that a stripe is an instability in a strongly correlated system, and$-$like the Wigner crystal instability in an electron gas \cite{wigner}$-$need not satisfy a nesting condition.  Hence, the word ``stripe" can be considered a euphemism for a charge/spin density wave in the strong coupling limit of a doped Mott insulator.

The original experimental evidence for a stripe smectic in the copper-oxides comes from neutron scattering, dating back to the original work on La$_{1.6-x}$Nd$_{0.4}$Sr$_x$CuO$_4$ by Tranquada.\cite{jtranNature}  This study established the existence of coexisting static spin and charge order whose wave vectors differed by a factor of two, implying a real space structure in which charges act as domain walls between antiferromagnetic regions.\cite{kivelsonNature}  In this phase the superconducting $T_c$ is suppressed, suggesting that stripe order is incompatible with superconductivity.\cite{moodenbaugh,jtranLNSCO1997}  However, subsequent inelastic neutron experiments established that, even in the superconducting state, stripe correlations may be observed at finite frequency, leading to speculation that dynamic stripes may enhance or even mediate superconductivity.\cite{kivelsonNature}  Regardless of whether stripes and superconductivity are directly related, it is important to understand this density wave phenomenon for the insights it gives into the cuprate phase diagram.

There are two important, outstanding questions left open by neutron experiments.  The first is whether the density wave observed is in fact a strong coupling effect and not a conventional SDW.  Because a neutron does not couple to charge, the so-called ``charge" peaks actually signify a distortion in the structural lattice, and do not necessarily imply the existence of charge order.  In fact, such a ``$2q$" structural modulation is always observed in conventional SDW materials, including Cr, because of magnetoelastic coupling between the spins and the lattice.\cite{hillCr,gruner}  Second, neutrons do not necessarily establish this phase to be a strong coupling effect, since the wave vector of the charge order is enticingly close to the $q_1$ nesting vector connecting the antinodal regions of the Fermi surface.\cite{hoffmanQI} Differentiating this phase from a simple SDW requires an experimental probe that is directly sensitive to charge order, as well as aspects of the electron spectral function that might provide insight into the strength of the interactions that drive it.

In this article we review efforts to use resonant soft x-ray scattering (RSXS) to study the electronic structure of density wave instabilities in cuprate and related materials.  RSXS is an elastic, x-ray diffraction technique in which the photon energy is tuned to resonantly excite electrons into the valence band, providing (through the intermediate states) momentum-resolved information about the electronic structure.  In particular we will describe how RSXS has not only established the existence of charge order, but has provided the most explicit evidence yet that the density wave state in cuprates is a strong correlation effect, and hence is not (purely) a consequence of Fermi surface nesting.

A key concept that has been missing from this field is a quantitative way to relate RSXS experiments to more established probes of valence band physics, namely angle-resolved photoemission spectroscopy (ARPES) and scanning tunneling spectroscopy (STM).  ARPES and STM are usually understood as measuring the electron spectral function, i.e., the imaginary part of the one-electron Green's function.  By contrast, most authors have up to now interpreted RSXS in terms of concepts borrowed from traditional x-ray diffraction, i.e., atomic form factors and structure factors, whose definitions are based on the assumption of optical locality in scattering.\cite{lectureNotes}  While this assumption can be used to understand elementary phenomena like charge order, it ignores the propagation of the photoexcited, valence band electron, so provides no understanding of the relationship of RSXS to the spectral function or more established electronic structure probes.

To rectify this situation, in Sections 3-6 we take a different perspective and evaluate the RSXS cross section in a limit that emphasizes the nonlocality of the intermediate states.  We show that, in cases where the excitonic interaction between the valence band electron and the core hole may be neglected (such as at the oxygen $K$ edge), RSXS measures a spectral function very similar to those measured by other valence band probes.  We argue that RSXS, as a momentum-space probe that detects real space inhomogeneity in the spectral function, forms a natural bridge between ARPES and STM techniques.

Finally, in Section 7, we outline what we view as the medium-term future of the RSXS technique, at least as it applies to high T$_c$ superconductivity.  In particular, we will argue that high magnetic field studies, which may be used to study the electronic structure of the vortex lattice, might be used to clarify the nature of the high field state, and hence explain the anomalous periodicities observed in quantum oscillation experiments.\cite{QO1,QO2}

\section{RSXS Studies of the Smectic Stripe State}

Resonant soft x-ray scattering is an elastic \cite{noteElas} x-ray diffraction technique in which the photon energy is tuned to resonantly excite core electrons into the valence band.  RSXS can be thought of as a two-step optical process involving a dipole absorption event followed by an emission event, mediated by a set of virtual, intermediate states.  In this sense, RSXS is similar to resonant Raman scattering (which if carried out in the x-ray regime is called ``RIXS"), except that the process is entirely elastic, i.e., in the final state the system is left in its ground state.  Through the intermediate states, RSXS accesses the physics of the valence band.

While requiring very different technology, RSXS is a simple conceptual evolution of the resonant hard x-ray magnetic scattering techniques first used in the 1980s to study magnetic ordering in rare earth materials.\cite{gibbs1,gibbs2}  The only distinction between the two techniques is the energy range: transitions into the valence band in transition metal oxides lie in the soft x-ray range ($400 \, eV < \hbar\omega < 1 \, keV$), which required the development of special, vacuum scattering techniques.\cite{abba2002} 

Compared to the more established valence band probes of ARPES and STM, RSXS has the disadvantage of poorer energy resolution, which is limited by the radiative lifetime of the intermediate states ($\Gamma \sim 150$ meV).  However RSXS has several advantages: as a bulk probe, it can be applied to any material, even those containing buried interfaces,\cite{smadici1,smadici2} or whose surfaces do not cleave. RSXS can also be applied to very small samples and can be carried out in a high field environment -- a point to which we will return shortly.

The first use of RSXS in any system appears to be the resonant reflectivity study of an iron film by Chi-Chang Kao in 1990.\cite{kao1990}  Its first application to strongly correlated oxides was an isolated study of La$_2$CuO$_{4+\delta}$,\cite{abba2002} but its early use was mainly to study manganites, where it had been argued to be only technique capable of directly probing orbital order.\cite{castleton,elfimov}  Experiments were carried out on a variety of manganite systems, which were then interpreted using concepts borrowed from traditional x-ray diffraction, namely the atomic ``form factor", $f^{ij}(\omega)$, \cite{castleton,wilkins,staub} which can be related to the integrated intensity of a Bragg reflection via the ``structure factor," \cite{ashcroft}

\begin{equation}
S_{\bf G} = \sum_n f_n^{ij} \, e^{-i {\bf G} \cdot r_n}.
\end{equation}

\noindent
The concept of a form factor is based on the assumption of optical locality, i.e., that the excitations that propagate the polarization, $P$, are local, or in other words that the photoexcited electron does not propagate.\cite{lectureNotes}  Assuming atomic level locality, most authors took the approach of using atomic multiplet calculations -- which had been used successfully for many years to analyze x-ray absorption experiments -- to determine these form factors, and comparing them to the RSXS energy line shapes.\cite{wilkins,stojic,thomas}.  Despite very creative efforts, however, quantitative agreement with the RSXS line shapes in manganites was never achieved.  The reasons for this are not completely understood, but may be related to this very issue of nonlocality; If nonlocal effects are important, an atomic description may not be valid.  In extreme cases, the concept of a form factor itself may be ill-defined.  We will return to this point in Section III. 

The first application of RSXS to a phenomenon directly related to stripes were studies of the so-called ``spin ladder" material, Sr$_{14-x}$Ca$_x$Cu$_{24}$O$_{41}$ (SCCO).\cite{abbaLadderNat}  The two-leg ladder was originally introduced by theorists as a computationally more tractable version of the two-dimensional $t-J$ model, which is still believed by many authors to contain some of the essential physics of high temperature superconductivity.\cite{dagotto1992}  Depending upon the parameters (e.g., the hole density), a doped ladder can exhibit either exchange-driven superconductivity \cite{dagotto1992,sigrist1994} or an insulating Wigner crystal ground state, in which the holes crystallize into a static lattice.\cite{dagotto1992,white2002}  The competition between these two phases is reminiscent of the competition, mentioned in Section I, between the ordered stripe smectic phase and superconductivity in two dimensions.\cite{jtranLNSCO1997}

The SCCO system contains both doped CuO$_2$ ladders and chains, and at various points in its phase diagram exhibits the transport characteristics of both of the above predicted ground states.  At $x=11$ SCCO exhibits (under hydrostatic pressure) superconductivity with $T_c=12.5 K$.\cite{uehara1996}  Further, at $x=0$ it exhibits clear transport characteristics of a charge density wave (CDW), for example a screening mode in impedance measurements, a nonlinear current-voltage ($I-V$) curve, etc.\cite{girsh2002,gorshunov2002}  Strangely, conventional x-ray scattering experiments, which are the standard way to study Peierls CDW materials such as NbSe$_3$ or K$_{0.3}$MoO$_3$ (``blue bronze"), revealed no superstructure in SCCO apart from the atomic supermodulation innate to this adaptive misfit material.\cite{fukuda,smaalen}  The reason for this absence was a mystery.

RSXS studies resolved this mystery by revealing that SCCO in fact contains a hole Wigner crystal.\cite{abbaLadderNat}  While conventional x-ray scattering is highly effective for studying Peierls CDWs, it does so by coupling to the lattice distortion rather than to the charge modulation itself.\cite{gruner}  Because a Peierls CDW is driven by the lattice distortion, the two always coexist and it is pedantic to debate which is being measured in the experiment.  In the case of a doped spin ladder, however, the charge modulation is driven by electron-electron interactions rather than a soft phonon, and a lattice distortion is not (in the first approximation) expected to occur.  For this reason, conventional x-ray techniques are not a natural way to study ordering in materials like SCCO.\cite{noteDistort}

Exploiting the valence band sensitivity of RSXS, studies of SCCO explicitly revealed the existence of a Wigner crystal in the ladder substructure.  This modulation was found to be stable only for commensurate hole filling, in agreement with predictions from density matrix renormalization group (DMRG) calculations.\cite{rusydiComm,white2002}  Further experiments also revealed that the CDW in the chain layer, for which evidence had been presented by several groups,\cite{matsuda,vuletic} was in fact due to interplay between hole ordering and the misfit supermodulation.\cite{rusydiUCR}  This last study exploited coherent interference between structural scattering and valence band scattering and demonstrated that phase-sensitive measurements could be used to objectively identify the character of a CDW.\cite{fDpaper}

A step closer to real stripe physics was achieved in a subsequent study of the nickelate material, La$_{1.8}$Sr$_{0.2}$NiO$_4$, which had been shown to exhibit simultaneous ordering of spins and holes, identifying it as a stripe material.\cite{jtranNickelate}  Performing RSXS experiments in resonance with the Ni $L$ edge, Sch\"u\ss ler-Langeheine and co-workers succeeded in measuring both spin and charge reflections, which both resonated strongly at the Ni $L$ edge, though only weakly at the La $M$ edge.\cite{schussler}  This established the order as residing in the NiO$_2$ planes, as expected.  In a first effort to account for nonlocal effects in scattering, the authors made use of the atomic form factors introduced in ref. \cite{castleton}, but computed their values using configuration-interaction, multiplet calculations on a cluster comprising not just a central Ni, but also its nearest neighbor oxygen ligands.  Complete charge disproportionation was assumed (i.e., all clusters were taken to have integer  Ni$^{2+}$ or Ni$^{3+}$ charge), and reasonably quantitative agreement with experiment was obtained.  

A direct connection to stripes and high $T_c$ superconductivity was finally achieved with a RSXS study of the stripe-ordered cuprate, La$_{2-x}$Ba$_x$CuO$_4$ (LBCO).\cite{LBCONatPhys}  This material has been cited as the prototypical stripe smectic, exhibiting the so-called ``1/8 anomaly" \cite{moodenbaugh} in which $T_c$ is suppressed at $x \sim 1/8$ due to the formation of quasi-long-ranged spin and charge order.\cite{jtranLBCO2004}  As discussed in Section I, while the existence of static spin order had clearly been established in this material by neutron scattering, the charge order was on less certain footing, since the observed charge peaks could also arise from magnetoelastic coupling between the spins and the lattice.  

RSXS studies, carried out at the oxygen $K$ edge, found that the charge reflections in this material exhibited a giant resonance at the Fermi energy, similar to that observed earlier in spin ladder materials, indicating the existence of pronounced, valence band charge order.\cite{LBCONatPhys}  The absolute scattering cross section was calibrated by scaling the intensity to the integrated weight of the (002) reflection of a reference crystal of Bi$_2$Sr$_2$CaCu$_2$O$_{8-\delta}$, whose structure factor is known.   Again applying the local scattering approximation, determining the form factors from doping-dependent x-ray absorption (XAS) data, the peak-to-trough oxygen valence amplitude was estimated to be $\Delta v=0.063$ holes.  Assuming a bond-centered pattern for the stripe order, \cite{lorenzana} this implied an integrated intensity of 0.59 holes under a single stripe, which was close to the value 1/2 expected for ideal, half-filled stripes.

In addition to the existence of charge order, this study revealed something unexpected: A second resonance was observed several eV above the Fermi energy, near the energy of the upper Hubbard band (UHB) (Fig. 1).  This observation implies not only a modulation in the doped hole density, but also a modulation in the amount of spectral weight in the UHB.  This indicates that the less doped, hole-poor, AF regions in the modulated stripe state are more Mott-like than the hole rich regions, or in other words that the degree of ``Mottness" of the system is itself modulated.\cite{LBCONatPhys}  This observation, which was confirmed by subsequent RSXS studies of stripe-ordered La$_{1.8-x}$Eu$_{0.2}$Sr$_x$CuO$_4$  \cite{fink1,fink2} and La$_{1.475}$Nd$_{0.4}$Sr$_{0.125}$CuO$_4$.\cite{wilkins2011}, implies that physics at the energy scale of the Hubbard $U$ plays a crucial role in the formation of the stripe state.

\begin{figure}
\includegraphics[width=1\textwidth]{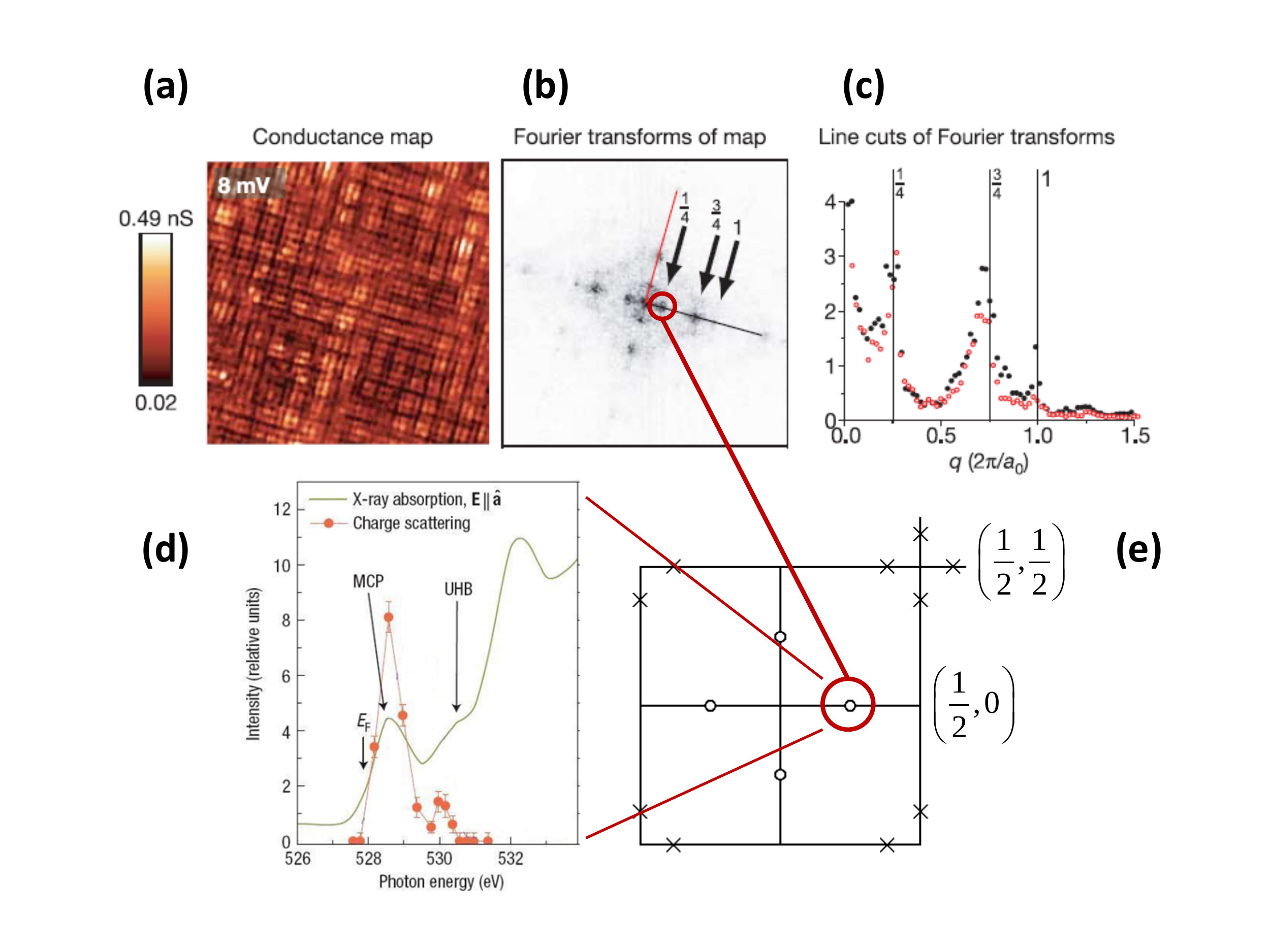}
\caption{Comparison between RSXS and Fourier transformed STM data, showing that the two measure essentially the same spectral function.  (a) STM spectroscopic map of the superconducting oxychloride Ca$_{2-x}$Na$_x$CuO$_2$Cl$_2$ (NCCOC), taken at a fixed 8 mV bias voltage, showing ``checkerboard" order.  (b-c) Fourier transform of the map in (a), showing that the density of states modulation has a wave vector of  $(k_x,k_y)=(1/4,0)$ in reciprocal lattice units. (d) RSXS energy scan with the momentum fixed at the same $(1/4,0)$ location in reciprocal space, this time in stripe-ordered  La$_{2-x}$Ba$_x$CuO$_4$, showing a density of states modulation with the same wave vector as in the oxychloride. Note the high energy peak near the upper hubbard band, which indicates a modulation in the degree of ``Mottness" in addition to charge order.  (e)  General cuprate Brillouin zone, showing locations of charge reflections (circles) and spin reflections (crosses). }
\end{figure}

In retrospect, the contribution of RSXS to the subject of stripe physics can be summarized as closing two issues that were left open by neutron scattering.  Specifically, RSXS has (1) established the existence of real charge order, and (2) provided the most explicit evidence yet that stripe order is -- at least in part -- a strong coupling effect.  This latter point does not prove that Fermi surface nesting is irrelevant to the formation of the stripe state, which again has a wave vector close to the nesting vector of the Fermi surface antinodes.  But it does suggest that the energy scale of the interactions that drive the instability is of order $U$.

Beyond spin ladders and 214 nickelates and cuprates, in which the order is static and long ranged, use of RSXS to study stripe physics has been limited.  Despite efforts by several groups, RSXS has failed to uncover charge order in any other cuprate, including Bi$_2$Sr$_2$CaCu$_2$O$_{8-\delta}$ (BSCCO), YBa$_2$Cu$_3$O$_{6+y}$ (YBCO), Ca$_{2-x}$Na$_x$CuO$_2$Cl$_2$ (NCCOC),\cite{serbanNCCOC} etc.  This has been particularly disappointing given that STM studies have produced extensive evidence for $4\times 4$ checkerboard order, as well as potentially related quasiparticle interference phenomena, in several of these systems.\cite{hanaguri,lawler}  Even studies of YBCO samples that seemingly exhibit a 1/8 anomaly\cite{liangAnomaly} showed no sign of charge order.  From these failures one is forced to conclude either that charge order is simply not a universal property of doped cuprates, or that it is present but is so disordered that the scattering is too diffuse to be distinguished from the inelastic background, which is rather high in RSXS.  This latter possibility has touched off debates over whether cuprates can be said to exhibit charge order when they are in fact disordered, and also whether a glassy, inhomogeneous background might in fact be more optimal for superconductivity.\cite{kivelsonOptimal,Okamoto,Berg}

\section{The Electron Green's Function}

We now return to the issue of nonlocality, i.e., the question of how one may interpret RSXS data when the photoexcited electron is not localized.  In the limit of extreme nonlocality an atomic or even cluster picture fails, and the concept of an atomic form factor, $f^{ij}$, is invalid.  The point we will attempt to make in this section is that, in this limit,  RSXS is better understood as measuring an electron spectral function, similar to those measured with ARPES or STM (though closer to the latter).  Specifically, the one-electron Green's function is defined at zero temperature as

\begin{equation}
G^{R,A}({{\bf r}},{{\bf r'}},t - t') = \pm i\left\langle 0 \right| \{ {\hat \psi ({{\bf r}},t),\,{{\hat \psi }^\dag }({{\bf r'}},t')} \} \left| 0 \right\rangle \theta[\pm(t-t')]) / \hbar.
\end{equation}

\noindent
where ``\{ , \}" denotes an anticommutator and the superscripts $R$ and $A$ denote a retarded or advanced Green's function and refer to the upper and lower signs, respectively.  We will find it convenient to separate $G$ into electron and hole components, i.e., 

\begin{equation}
G^{R,A}({{\bf r}},{{\bf r'}},t - t') = G^{R,A}_e({{\bf r}},{{\bf r'}},t - t') + G^{R,A}_h({{\bf r}},{{\bf r'}},t - t')
\end{equation}

\noindent where

\begin{equation}
G^{R,A}_e({{\bf r}},{{\bf r'}},t - t') = \pm i\left\langle 0 \right| \hat \psi ({\bf r},t) \,{{\hat \psi }^\dag }({\bf r'},t') \left| 0 \right\rangle \theta[\pm(t-t')]) / \hbar
\end{equation}

\noindent and 

\begin{equation}
G^{R,A}_h({{\bf r}},{{\bf r'}},t - t') = \pm i\left\langle 0 \right|  {{\hat \psi }^\dag }({{\bf r'}},t') \, \hat \psi ({\bf r},t) \left| 0 \right\rangle \theta[\pm(t-t')]) / \hbar .
\end{equation}

\noindent Crucially, we allow here for the possibility that the system may be inhomogeneous, i.e., translational symmetry is broken either explicitly by disorder or spontaneously by the formation of a density wave.  In this case, $G^{R,A}({\bf r},{\bf r}',t-t')$ is not a function only of ${\bf r} - {\bf r'}$, but is an independent function of ${\bf r}$ and ${\bf r}'$, and its Fourier transform, $G^{R,A}({\bf k},{\bf k}',\omega)$, is an independent function of ${\bf k}$ and ${\bf k}'$.  In the ensuing discussion we will also refer to the mixed representation, $G^{R,A}({\bf r},{\bf r}',\omega)$.

The beauty of ARPES and STM, which underlies the enormous impact these probes have had in condensed matter physics, is that they both measure aspects of the electron Green's function.  Assuming the sudden approximation and neglecting matrix element effects, ARPES measures the diagonal components of the hole spectral function in momentum space,\cite{campuzano}

\begin{equation}
A_{ARPES}({\bf k},\omega)=A_{h}({\bf k},-{\bf k},\omega) = \mp \frac{1}{\pi} Im \left [ \, G^{R,A}_h( {\bf k}, -{\bf k},\omega) \, \right ],
\end{equation}

\noindent and STM measures the diagonal components of the full spectral function in real space,

\begin{equation}
A_{STM}({\bf r},\omega) = A({\bf r},{\bf r},\omega) = \mp \frac{1}{\pi} Im  \left [ \, G^{R,A} ( {\bf r}, {\bf r},\omega) \, \right ],
\end{equation}

\noindent
which is basically the local density of states.  Note that eqs. 2-5 imply that $A_e(\omega)=\theta(\omega) A(\omega)$ and $A_h(\omega)= [1-\theta(\omega)] A(\omega)$, i.e., $A_e(\omega)+A_h(\omega)=A_e(\omega)$.  We emphasize that, because ARPES and STM probe only the imaginary part, it is unimportant whether a retarded or advanced causality convention is used, since both give$-$modulo an overall sign$-$the same result in eqs. 6 and 7.  

Framed in this manner, it is clear that ARPES and STM data cannot be directly related to one another by a spatial Fourier transform.  While both probes measure a spectral function, their relationship involves off-diagonal terms that are not measured in either experiment.  Hence, relating the two requires a model of quasiparticle scattering,\cite{seamusQI} perhaps involving an autocorrelation approach.\cite{jcc}

\section{Resonant Scattering}
We now wish to determine what spectral function would be measured with RSXS assuming the nonlocal limit of the intermediate states.  We will show that, despite being a momentum-space probe, RSXS is more closely related to STM measurements than to ARPES measurements, its spectral function bearing a close resemblance to eq. 7.  We begin by coupling light to the electrons by defining the classical canonical momentum, 

\begin{equation}
{\bf p} \rightarrow {\bf p} + \frac{e}{c} \hat{\bf A}
\end{equation} 

\noindent
where ${\bf A}$ is the field operator for the photons.  This coupling yields two interaction Hamiltonians between the electrons and photons,\cite{sinha}

\begin{equation}
\hat{H}_1 = \int{ d{\bf r} \, \hat{\psi}^\dagger \frac{e^2 {\bf A}^2 }{2mc^2}  \hat{\psi} }.
\end{equation}

\noindent 
and

\begin{equation}
\hat{H}_2 = \int{ d{\bf r} \, \hat{\psi}^\dagger \frac{e {\bf A} \cdot {\bf p} }{2mc}  \hat{\psi} },
\end{equation}

\noindent
where we have taken $\nabla \cdot {\bf A} = 0$.  The amplitude 

\begin{equation}
S_{f \leftarrow i} = \left\langle f \right|U\left( {\infty , - \infty } \right)\left| i \right\rangle
\end{equation}

\noindent
produces scattering through $\hat H_1$ to first order in perturbation theory. This term is the origin of Thomson scattering and conventional, nonresonant inelastic x-ray scattering.\cite{sinha,advmat}  Resonant scattering arises through $\hat H_2$ acting in second order,

\begin{equation}
S_{f \leftarrow i} = \frac{2 \pi}{\hbar} \sum_n{ \frac{\langle f | \hat H_2(0) | n \rangle \langle n | \hat H_2(0) | i \rangle }{ E_i - E_n - i\eta} }.
\end{equation}

\noindent
where $\eta$ is the radiative lifetime of the intermediate states.  Here the initial and final states both correspond to the system in its ground state in the presence of a photon, i.e.,

\begin{equation}
|i\rangle = a^\dagger_{{\bf k}_i \epsilon_i} |0\rangle \nonumber
\end{equation}
and
\begin{equation}
|f\rangle = a^\dagger_{{\bf k}_f \epsilon_f} |0\rangle
\end{equation}
where the vacuum $|0 \rangle$ represents the ground state of the many-body electron system, including the core electrons, and the intermediate states $|n \rangle$ describe all the many-body excitations.  ${\bf k}_i$ and ${\bf k}_f$ represent the initial and final state of the photon, and ${\epsilon}_i$ and ${\epsilon}_f$ are the corresponding polarizations.

\section{Lattice Model: RSXS Amplitude}

For illustrative purposes, we consider now the case where our many-body system resides within a single band on a discrete lattice.  While this is a significant simplification, we anticipate that the conclusions we draw will be valid quite generally.  Note that we do {\it not} assume that the interaction among the valence electrons is weak or that any particular low energy theory necessarily applies.  

Within this model, our goal is to rewrite eq. 12 so that it resembles a spectral function for the valence electrons.  The electron field operator is

\begin{equation}
\hat \psi ({\bf{r}}) = \frac{1}{{\sqrt 2N }}\sum\limits_j {{d_j}{\varphi _j}({\bf{r}}) + {c_j}\varphi _j^c({\bf{r}})}
\end{equation}

\noindent
where $\phi_j({\bf r})=\phi({\bf r}-{\bf r}_j)$ represents the single-particle basis function for a $d$ orbital residing on site $j$, and $\phi^c_j({\bf r})$ is the core orbital on the same site.  With this reduced basis set, the interaction in eq. 10 has the simplified form

\begin{equation}
\hat H_2 =  \sum\limits_{j,{\bf{k}},\epsilon_{\bf k} } {V\left( {{\bf{k}},{{\epsilon}_{{\bf{k}}}}} \right) \left ( d_j^\dag a_{{\bf{k}}\epsilon_{\bf k}} c_j \, e^{i{\bf{k}} \cdot {{\bf{r}}_j}} \right . } \\
+ \left . c_j^*a_{{\bf{k}}\epsilon_{\bf k} }^\dag d_j \,{e^{ - i{\bf{k}} \cdot {{\bf{r}}_j}}} \right )
\end{equation}

\noindent 
where $a_{{\bf k}\epsilon_{\bf k}}$ annihilates a photon and the potential

\begin{equation}
V\left( {{\bf{k}},\epsilon _{{\bf{k}}}} \right) = \frac{e}{{mN}}\sqrt {\frac{{\pi \hbar }}{{2V{\omega _k}}}} F(\epsilon _{{\bf{k}}})
\end{equation}

\noindent depends on the polarization, $\epsilon_{\bf k}$, through the dipole matrix element

\begin{equation}
F(\epsilon_{\bf k} ) = \int{ d{\bf r} \, \phi^*({\bf r}) \, \hat \epsilon_{\bf k} \cdot {\bf p} \, \phi^c({\bf r}) }.
\end{equation}

\noindent
The numerator of eq. 12 then takes the form

\begin{equation}
\left\langle f \right|{\hat H_1}\left( 0 \right) \left| n \right\rangle \left\langle n \right|{\hat H_1}\left( 0 \right)\left| i \right\rangle  = V\left( {{{\bf{k}}_f},{{\hat \epsilon }_{{{\bf{k}}_f}}}} \right)V\left( {{{\bf{k}}_i},{{\hat \epsilon }_{{{\bf{k}}_i}}}} \right) \sum\limits_j {{e^{-i \bf{q} \cdot {{\bf{r}}_j}}}\left\langle 0 \right|\,c_j^*d_j^{}\left| n \right\rangle \left\langle n \right|d_j^\dag {c_j}\,\left| 0 \right\rangle }
\end{equation}

\noindent
where

\begin{equation}
\bf{q}=\bf{k}_f - \bf{k}_i
\end{equation}
is the momentum transfer.

To manipulate this into a form that resembles a spectral function for the valence electrons, it is necessary to eliminate the core operators, $c_j$ and $c_j^\dagger$.  We wish to do this in a way that preserves the full nonlocality of the intermediate states, $|n>$.  To this end, we make the assumption that the excitonic interaction between the core hole and the photoexcited valence electron is negligibly small.  This nonlocal limit is the diametric opposite of that usually used to justify the atomic multiplet approach.\cite{carraThole}   In this approximation, the intermediate states factor into core and valence components, i.e.,

\begin{equation}
|n>=|\bar{c}>|n^{N+1}>.
\end{equation}

\noindent
Here $|\bar{c}>$ represents a hole in a core level and $|n^{N+1}>$ represents the $n$th excited state of the valence electron system containing $N+1$ electrons, one having been promoted from the core.  

Is this approximation valid?  In most cases, it certainly is not.  It should be expected to fail dramatically, for instance, at transition metal $L$ edges, where excitonic core hole effects are dominant and nonlocal effects can sometimes be ignored entirely. This approximation may be reasonable, however, at some ligand K edges, such as the O $K$ edge.  Further, as discussed earlier, nonlocal effects may be important even at certain $L$ edges, such as Mn.  So it is important to consider, as a point of reference, what form the cross section has in the nonlocal limit.

Assuming eq. 21 holds, the contractions in eq. 18 reduce to

\begin{equation}
<n| d_j^\dagger c_j |0> = <n^{N+1}| d_j^\dagger |0> <\bar{c}| c_j |0> \\
= <n^{N+1}| d_j^\dagger |0>
\end{equation}

It is also helpful to rewrite the energy denominator in eq. 12 in terms of familiar, microscopic quantities.  The initial energy can be expressed as

\begin{equation}
E_i = \hbar \omega + E_0^N + \epsilon_{core}
\end{equation}

\noindent
where $\hbar\omega$ is the photon energy, $E_0^N$ is the ground state energy of the $N$ valence electron system, and $\epsilon_{core}$ is the energy of the core electron.  Recognizing that the intermediate state energy is just one of the excitation energies of the $N+1$ electron system,

\begin{equation}
E_n = E_n^{N+1},
\end{equation}

\noindent
we can rewrite the energy denominator in eq. 10 as

\begin{equation}
E_i-E_n =  \hbar \omega + E_0^N - E_n^{N+1} + \epsilon_{core}.
\end{equation}

\noindent
It is even more convenient to express this difference in terms of the chemical potential,

\begin{equation}
\mu  \equiv E_0^{N + 1} - E_0^N,
\end{equation}

\noindent
and the excitation energy (which might be a quasiparticle energy) of the $N+1$ valence electron system, 

\begin{equation}
\epsilon _n^{N+1} = E_n^{N + 1} - E_0^{N + 1}.
\end{equation}

\noindent
In terms of these quantities, the final form for the energy denominator is 

\begin{equation}
E_i-E_n = \hbar\omega - \epsilon_n^{N+1} - \mu + \epsilon_{core}.
\end{equation}

\noindent
The scattering amplitude then has the form

\begin{equation}
S_{f \leftarrow i} = \frac{2\pi}{\hbar} V\left( {{{\bf{k}}_f},{{\hat \varepsilon }_{{{\bf{k}}_f}{\alpha _f}}}} \right)V\left( {{{\bf{k}}_i},{{\hat \varepsilon }_{{{\bf{k}}_i}{\alpha _i}}}} \right) \sum_{n,j}{  \frac{ \left | <n^{N+1}|d_j^\dagger|0> \right |^2 }{\hbar\omega - \epsilon_n^{N+1} - \mu + \epsilon_{core} -i\eta} } e^{-i {\bf q} \cdot {\it r}_j}
\end{equation}

\noindent
We will now perform an approximation that is rarely done in x-ray disciplines, though is done very commonly by ARPES and STM groups, which is to ignore the matrix elements, i.e., 

\begin{equation}
V\left( {{{\bf{k}}_f},{{\hat \varepsilon }_{{{\bf{k}}_f}{\alpha _f}}}} \right)V\left( {{{\bf{k}}_i},{{\hat \varepsilon }_{{{\bf{k}}_i}{\alpha _i}}}} \right)\approx 1
\end{equation}

\noindent
This approximation is nonstandard but is probably valid for measurements over a narrow energy range at approximately constant geometry.  The resulting expression for the RSXS scattering amplitude is

\begin{equation}
{S_{f\leftarrow i}}({\bf{q}},\omega ) = {\sum\limits_{n,j} {\frac{{{{\left| {\left\langle {n{}^{N + 1}} \right|d_j^\dag \,\left| 0 \right\rangle } \right|}^2}}}{{\hbar \omega  - \epsilon _n - \mu  + {\epsilon _{core}} - i\eta }}{e^{ - i{\bf{q}} \cdot {{\bf{r}}_j}}}} } ,
\end{equation}
which looks encouragingly like a spectral function.  The scattered intensity $I_{RSXS}({\bf q},\omega)= \left | S_{f \leftarrow i}({\bf q},\omega) \right |^2$.  To make a precise statement, however, an explicit comparison is needed.

\section{Lattice Model: Spectral Functions for STM and RSXS}

To make this comparison, we evaluate the spectral functions introduced in Section 3 for this lattice model. We will focus on the form suitable for STM, which as mentioned earlier will turn out to be the closer relative of RSXS.  In the lattice model, the one electron Green's function has the form

\begin{equation}
G^{R,A}(j,j',t-t') =  \pm i \left\langle 0 \right|\left\{ {{d_j}(t),d_{j'}^\dag (t')} \right\}\left| 0 \right\rangle \theta \left[ \pm(t - t') \right ] / \hbar
\end{equation}
where $j$ is a site index.  Written out explicitly,

\begin{equation}
G^{R,A}(j,j',\omega ) = G^{R,A}_e(j,j',\omega ) + G^{R,A}_h(j,j',\omega )
\end{equation}

\noindent where

\begin{equation}
G^{R,A}_e(j,j',\omega ) =  \sum\limits_n \frac{{\left\langle 0 \right|{d_j}\left| n^{N + 1} \right\rangle \left\langle {n^{N + 1}} \right|d_{j'}^\dag \left| 0 \right\rangle }}{\hbar \omega  - \epsilon _n^{N+1} \pm  i\gamma }
\end{equation}

\noindent and

\begin{equation}
G^{R,A}_h(j,j',\omega ) =  \sum\limits_n \frac{{\left\langle 0 \right|d_j^\dag \left| {n^{N - 1}} \right\rangle \left\langle {n^{N -1}} \right| d_{j'} \left| 0 \right\rangle }}{\hbar \omega  + \epsilon _n^{N-1} \pm  i\gamma }
\end{equation}

\noindent  Neglecting matrix element effects, the STM spectrum is given by the diagonal spectral function

\begin{equation}
A_{STM}(j,\omega ) =  \mp \frac{1}{\pi }{\mathop{\rm Im}\nolimits} \left[ {G^{R,A}(j,j,\omega )} \right]
\end{equation}

\noindent which, again modulo a sign, should be the same whether a retarded or advanced causality convention is chosen. 

Returning to the subject of RSXS, we now consider the real space diagonal values of the {\it advanced} electron Green's function, which have the form

\begin{equation}
G^A_e(j,j,\omega ) = \sum_n \frac{\left |  \left <n^{N+1} \right |  d^\dagger_j \left | 0 \right >  \right |^2 }{\hbar \omega  - \epsilon _n^{N+1} -  i\gamma}.
\end{equation}

\noindent Taking the real space Fourier transform of this quantity gives

\begin{equation}
\tilde G^A_e({\bf q},\omega ) = \sum\limits_j {G^A_e(j,j,\omega )\, e^{ - i \bf{q} \cdot {\bf r}_j}} = \sum_{n,j} \frac{\left |  \left <n^{N+1} \right |  d^\dagger_j \left | 0 \right >  \right |^2 }{\hbar \omega  - \epsilon _n^{N+1} -  i\gamma}  e^{ - i \bf{q} \cdot {\bf r}_j}.
\end{equation}

\noindent
A tilde has been added to distinguish this quantity from the momentum space function $G^R_h({\bf k},-{\bf k},\omega)$ measured with ARPES, defined in eq. 6.  

We now perform a final manipulation on eq. 31, which is to write it in terms of an offset energy, $\hbar \omega' = \hbar \omega - \mu + \epsilon_{core}$, which is the photon energy measured with respect to the chemical potential, $\mu$.  This gives

\begin{equation}
S'_{f\leftarrow i}({\bf{q}},\omega')  \equiv S_{f\leftarrow i}({\bf{q}},\omega) = {\sum\limits_{n,j} {\frac{{{{\left| {\left\langle {n{}^{N + 1}} \right|d_j^\dag \,\left| 0 \right\rangle } \right|}^2}}}{{\hbar \omega ' - \epsilon _n^{N+1} - i\eta }}{e^{ - i{\bf{q}} \cdot {{\bf{r}}_j}}}} }.
\end{equation}

\noindent The scattered intensity in terms of this energy is then $I'_{RSXS}({\bf q},\omega)=\left | S'_{f \leftarrow i}({\bf q},\omega)  \right |^2 $ .   By inspection, one can see that eq. 39 is identical to the advanced Green's function in eq. 38, except in one respect: the damping $\gamma$ is infinitesimal, while the radiative lifetime $\eta$ is finite.  

As mentioned in Section 2, the lifetime issue is essentially one of resolution:  The scattering amplitude in eq. 39 can be thought of as the Green's function in eq. 38 convolved with a Lorentzian with width $\eta$.  Spectral features on a scale finer than $\eta$ will be washed out.
Once one accepts resolution limitations, however, the following relationship can be considered to hold:

\begin{equation}
I'_{RSXS}({\bf q},\omega) = \left | G^A_e({\bf q},\omega ) \right |^2.
\end{equation}

Hence, we conclude that$-$in the extreme nonlocal limit$-$RSXS measures the square modulus of the Fourier transform of an advanced version of Green's function measured in STM experiments.  Extrapolating this result beyond our lattice model, we arrive at a tentative but direct relationship between STM and RSXS measurements:

\begin{equation}
I'_{RSXS}({\bf q},\omega) \sim \left | \int d{\bf r} \int_0^\infty d\omega' \frac{A_{STM}({\bf r},\omega)}{\omega-\omega'-i\eta} e^{-i{\bf q}\cdot {\bf r}} \right |^2
\end{equation}

\noindent 
Despire the resolution limitations, from this relationship we see a crucial advantage of RSXS over the more common probes of STM and ARPES:  RSXS is sensitive to the real part of the Green's function, rather than just its imaginary part.  This has several implications.  First, RSXS is sensitive to the causality convention chosen, and in particular requires use of the {\it advanced} $G$ if comparison to a microscopic model is performed.  Second, RSXS is capable of certain classes of phase-sensitive, interference experiments.  This has already been exploited to measure the phase shift between the charge and strain waves in the CuO$_2$ chain layer in Sr$_{14}$Cu$_{24}$O$_{41}$.\cite{rusydiUCR}

The similarity of RSXS to STM, rather than ARPES, is intuitive if one considers that the former two techniques probe the static inhomogeneity of a system.  For example, suppose one had a system that was perfectly homogeneous.  An ARPES measurement would yield a highly featured spectral function, $A({\bf k},\omega)$, whose momentum structure is reflective of the propagation characteristics of the quasiparticles.  An STM measurement, however, would be trivial:  The same  spectrum would be acquired at every location, the real space spectral function, $A({\bf r},\omega)$, being independent of {\bf r}.  Similarly, there being no real space structure to diffract from, a RSXS measurement would only observe scattering in the forward (${\bf q}=0$) direction.  In an inhomogeneous system, however, the real space spectral function, $A({\bf r},\omega)$, would acquire a non-trivial dependence on {\bf r}, and both STM and RSXS measurements would yield non-trivial results.  

\section{Outlook for the RSXS Technique}

We now try to provide some perspective on the future of the RSXS technique.  Because of its pronounced inelastic background, RSXS has proven to be of rather limited use for studying systems with short-ranged or glassy order, such as NCCOC.\cite{hanaguri}  Attempts to remove this inelastic background with multilayer analyzers have not been fruitful because of problems controlling geometric contributions to the energy resolution.  Other possible routes to energy analysis, such as transmission gratings or backscattering crystals, may be more fruitful but require significant instrumentation development that has only just begun.  Hence, it is useful consider whether there might be a more natural direction for the RSXS technique.

The core strengths of RSXS as a valence band probe are (1) naturally high momentum resolution, i.e., RSXS can detect structures $\sim 10^2$ nm in size without any special efforts, (2) bulk sensitivity, and (3) compatibility with a high field environment, where for example ARPES methods fail.  This suggests that RSXS is most naturally suited to large, highly ordered  phenomena that have a nontrivial magnetic field dependence.  

These strengths imply two directions.  The first is artificially structured transition metal oxides, for example superlattices created with layer-by-layer synthesis techniques such as molecular beam epitaxy (MBE) or pulsed laser deposition (PLD), or thin films that have been lithographically patterned.  The former are of tremendous current interest in the field of oxide interfaces, in which new emergent phenomena such as superconductivity and magnetism have been observed.\cite{mannhart}  RSXS is a natural probe of such interfaces, and has already been used to detect an emergent Fermi surface at LaMnO$_3$-SrMnO$_3$ interfaces \cite{smadici1} and charge delocalization in heterostructures of La$_2$CuO$_4$ and La$_{1.64}$Sr$_{0.36}$CuO$_4$\cite{smadici2}.  Preliminary studies of large arrays of La$_{1-x}$Sr$_x$MnO$_3$ quantum wires also seem promising.\cite{chenWires}

The second direction is high magnetic field studies.  Several groundbreaking quantum oscillations experiments \cite{QO1,QO2} have, for the first time, demonstrated the existence of well-defined quasiparticles in copper-oxide superconductors.  These measurements, which are performed in magnetic fields higher than 60 T, are the first explicit evidence for the validity of the quasiparticle concept, which places strong constraints on the types of theories that apply to cuprate materials.  Unfortunately, the oscillation periods observed in these measurements do not seem to correspond to the volume of any Fermi surface pocket observed with ARPES techniques.  The leading explanation of this apparent contradiction is that the cuprates form an ordered state at high temperature -- perhaps even a stripe smectic which some measurements already suggest is stabilized at high fields\cite{hoffman,lake,julien} -- whose correlation length is long enough to fold the Fermi surface, creating new pockets.  Hence, clarifying the validity of the quasiparticle concept in cuprates seems to boil down to the problem of determining whether an ordered state exists at high fields, and what it's origin might be.  This is a natural problem for RSXS, which is kinematically well matched to the length scales of the vortex lattice, and most likely whatever valence band instability causes the Fermi surface folding.  Development of high-field instruments, which will be a first for the RSXS texcnique, is already underway at the SSRL facility in Menlo Park, CA and the BESSY facility in Berlin.

\section{Acknowledgments}
We thank Eduardo Fradkin for many helpful discussions.  This work was supported by the Center for Emergent Superconductivity, an Energy Frontier Research Center funded by the U.S. Department of Energy, Office of Science, Office of Basic Energy Sciences under Award Number DE-AC0298CH1088. The derivation of the spectral function was supported by DOE grant DE-FG02-06ER46285.





\bibliographystyle{elsarticle-num}
\bibliography{stripe-review-spectral}

\begin{thebibliography}{10}
\bibitem{noteElas}
To be more precise, RSXS is an inergy-integrated measurement. As currently
  practiced, RSXS measurements employ no energy analyzer on the scattered
  radiation, so technically detect both elastic and inelastic processes. That
  said, most studies so far have focused phenomena exhibiting long ranged
  order, in which case the scattering is concentrated into localized regions of
  reciprocal space where the elastic signal dominates.
\bibitem{noteDistort}
A skeptic might argue that even a Wigner crystal must, by symmetry, exhibit
  some lattice distortion, since the lattice should respond to the charge
  modulation. In principle this argument is correct, but in practice there is
  an important quantitative distinction: While the lattice distortion in a
  Peierls material such as NbSe$_3$ is $\sim 0.05 \AA$, the distortion in
  Sr$_{14}$Cu$_{24}$O$_{41}$ is of order $10^{-3}\AA$, i.e., more than an order
  of magnitude smaller, and the scattered intensity will be smaller by the
  square of this ratio.
\end{thebibliography}







\end{document}